\documentclass[sigconf]{acmart}
\copyrightyear{2026}
\acmYear{2026}
\setcopyright{cc}
\setcctype{by}
\acmConference[SIGIR '26] {Proceedings of the 49th International ACM SIGIR Conference on Research and Development in Information Retrieval}{July 20--24, 2026}{Melbourne, VIC, Australia.}
\acmBooktitle{Proceedings of the 49th International ACM SIGIR Conference on Research and Development in Information Retrieval (SIGIR '26), July 20--24, 2026, Melbourne, VIC, Australia}
\acmISBN{979-8-4007-2599-9/2026/07}
\acmDOI{10.1145/3805712.3808419}

\usepackage{graphicx}
\usepackage{epstopdf}
\begin{document}

\title{Cheaper is Better: A Discount-Aware Network for Conversion Rate Prediction in E-commerce Recommendation System}


\author{Ruocong Tang}
\affiliation{%
  \institution{Alibaba Group}
  \city{Hangzhou, 311121}
  \country{China}}
\email{tangruocong.trc@taobao.com}

\author{Yang Huang}
\authornote{Corresponding author.}
\affiliation{%
  \institution{Alibaba Group}
  \city{Hangzhou, 311121}
  \country{China}}
\email{hy234680@taobao.com}

\author{Xing Fang}
\affiliation{%
  \institution{Alibaba Group}
  \city{Hangzhou, 311121}
  \country{China}}
\email{fangxing.fx@taobao.com}

\author{Chenyi Yan}
\affiliation{%
  \institution{Alibaba Group}
  \city{Hangzhou, 311121}
  \country{China}}
\email{yanchenyi.ycy@taobao.com}

\author{Chuike Sun}
\affiliation{%
  \institution{Alibaba Group}
  \city{Hangzhou, 311121}
  \country{China}}
\email{sunchuike.sck@taobao.com}

\author{Jing Wang}
\affiliation{%
  \institution{Alibaba Group}
  \city{Hangzhou, 311121}
  \country{China}}
\email{jing.wangj1@taobao.com}

\renewcommand{\shortauthors}{Ruocong Tang et al.}

\begin{abstract}
Post-click conversion rate (CVR) is a crucial element in online recommendation systems, which addresses significant challenges such as data sparsity (DS), sample selection bias (SSB), and delayed feedback. However, the impact of item discount rate-a key factor influencing both pricing and user purchasing behavior, has received limited attention. In this paper, we introduce the Discount-Aware Network (DANet) to model the relationship between item discount rates and CVR. DANet comprises three main components: 1) a time-frequency transformation module that utilizes Fourier transform to derive the frequency spectrum and capture the long-term discount rate trends of items; 2) a distribution de-bias module designed to mitigate the biases in user-specific discount rates caused by various purchase combinations and promotional activities, as well as periodic deviations linked to different promotion periods on e-commerce platforms; and 3) a supervised regression auxiliary task that establishes the explicit item discount labels to enhance the model's performance in terms of value accuracy, facilitating an effective representation of item discount rates. Experimental results on real datasets demonstrate the superiority of DANet, with offline AUC improving by 1.61\%, and online A/B test also shows that DANet achieves impressive gains of 3.63\% on pCVR and 2.23\% on GMV. DANet has been successfully deployed on Alibaba Tmall APP. The code is available at \url{https://github.com/tangrc/DANet}.
\end{abstract}

\ccsdesc[500]{Information systems~Recommender systems}

\keywords{Recommender System; Conversion Rate Prediction; Discount Rate}


\maketitle
\vspace{-0.4cm}  

\section{Introduction}
\label{sec:intro}

\begin{figure}[ht]
\centering
\includegraphics[scale=0.34]{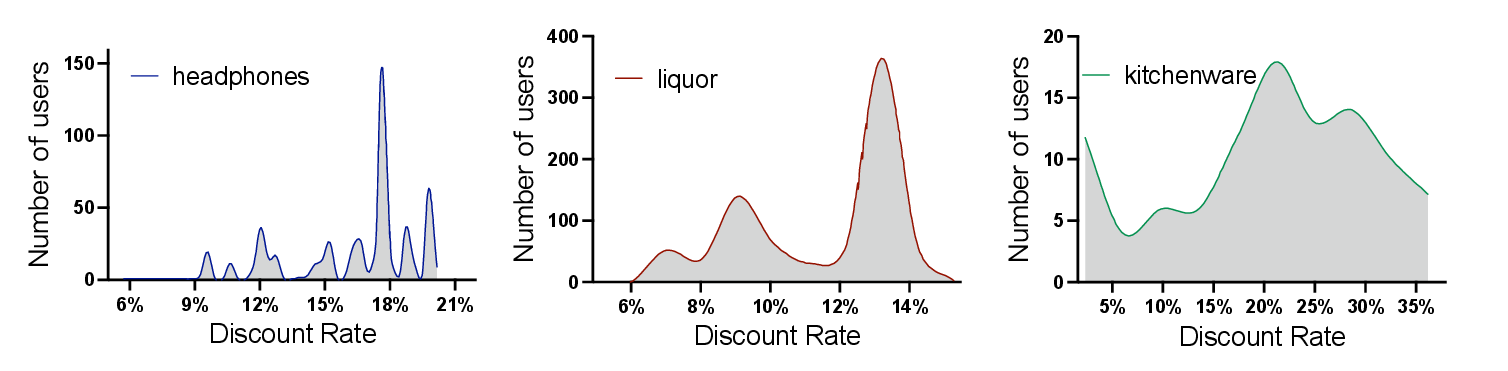}
\caption{Examples of varying discount rates for actual purchases due to different levels of user participation in promotional activities, despite the platform offering the same price for a specific item to all users.}
\label{figure1}
\end{figure}

With the rapid development of the Internet, various e-commerce platforms 
have invested significant resources in developing 
recommendation systems to enhance personalized item recommendations and increase gross merchandise volume (GMV)~\cite{li2019multi,pi2020search,ma2018modeling,Pan_2022}. This paper mainly focuses on the conversion rate (CVR) prediction problem. Researchers and practitioners have proposed a variety of methods to improve the accuracy of CVR model from many aspects, such as data sparsity~\cite{lyu2023entirespacelearningframework}, sample selection bias~\cite{huang2024utilizing, zhu2022evolutionpopularitybiasempirical} , and delayed transactions~\cite{chen2022asymptoticallyunbiasedestimationdelayed}.

Despite the previous advancement,  we notice that the impact of price on CVR remains underexplored. \cite{feng2023modeling} introduces item price as a new factor in CVR prediction and simulates user price tolerance. \cite{Zhang_2022} is proposed to jointly model the user interest and price preferences, along with the balance between them. As a pivotal factor influencing consumer purchasing decisions, the complexity and variability of price make direct modeling of price a formidable challenge~\cite{umberto2015developing}. Specifically, item prices could range from as low as a few yuan to tens of thousands, with significant distribution variations across different sellers and item categories. These variations present substantial obstacles to effectively learning the relationship between price and CVR.

 To address this issue, we propose using $Discount\ Rate (DR)$ as a substitute for price. DR effectively captures item price changes and reveals underlying marketing trends on the platform. Numerous studies have demonstrated that DR has a significant effect on influencing the consumer behaviors and purchasing decisions ~\cite{zhao2015commerce,qibtiyah2021influence,lan2021new,gong2018discounts,li2024commerce}.  Furthermore, we argue that DR is influenced by a triadic interplay of factors: 1) long-term pricing strategies by sellers for different item categories, 2) promotional discount activities organized by the platform, and 3) the level of user sensitivity to discounts and their participation in these discount activities. Figure ~\ref{figure1} illustrates the distribution of user-purchased DR for items such as headphones, liquor, and kitchenware over a 90-day period,  which reflects that different users have different sensitivities to the DR of the same item. Notably, the dynamic and personalized nature of discount activities (e.g., spend-and-save incentives) makes DR difficult to acquire in advance until the actual payment amount is determined at checkout.

Based on the above observations, we propose a novel CVR model: Discount-Aware Network (DANet), which comprehensively perceives the long-term DR trends of items, the context related to DR (such as platform promotions and item categories), and user sensitivities to DR, thereby achieving more accurate CVR predictions. First, we design a time-frequency transformation module, where Fast Fourier Transform (FFT)~\cite{soliman1990continuous} is applied to the time series to obtain its spectrum, containing comprehensive information about the time-frequency distribution of item DR. Second, we constructed a distribution correction module to correct DR sensitivity deviations at the user level and periodic DR deviations caused by different promotion periods. Finally, a supervised regression auxiliary task is designed to utilize explicit item DR labels, assisting network learning in obtaining a well-represented DR representation. The main contributions of our work are summarized as follows:

\begin{itemize}
    \item To address the problem that the CVR model cannot perceive the long-term price trends of items, we introduce a modeling method of personalized $Discount\ Rate(DR)$ in CVR prediction which achieve high prediction accuracy by self-adapting to different periods (e.g. flat sale period and promotion period).
    \item We propose a novel \textbf{D}iscount-\textbf{A}ware \textbf{N}etwork(DANet), which includes a time-frequency transformation module, a distribution correction module, and a supervised regression auxiliary task, allowing for a more comprehensive understanding of long-term DR trends of item and users' sensitivity to DR.
    \item Experiments on real datasets demonstrate that our DANet model outperforms representative methods. \textbf{DANet has been successfully deployed on Alibaba Tmall APP}(an e-commerce App with tens of millions of daily active users (DAU)), which implementation increases response time by less than 5\%, acceptable for our recommendation system, resulting in significant improvements in core performance metrics (pCVR +3.63\%, GMV +2.23\%).
\end{itemize}

\section{The Proposed Method}
\subsection{Problem Definition}
Given a training dataset $\mathcal{D}=\left\{\left(x_i, y_i\right)\right\}_{i=1}^{|\mathcal{D}|}$, $|\mathcal{D}|$ denotes a sample, where $x_i=\left\{{x}^{seq}, {x}^{u}, {x}^{i},{x}^{c},{S}^{d}\right\}$ denotes the high-dimensional feature vector, and $y_i$ is the CVR label, with $y_i=1$ indicating the sample is purchased. Our task is to predict the probability of CVR $p_{cvr}=p(y_i=1 \mid x_i)$for the test sample $x_i$. Specifically, features consist of five parts: 1) user behavior sequence $\boldsymbol{x}^{s e q}$, including click seq, cart seq, buy seq; 2)user features $\boldsymbol{x}^{u}$, including basic user profile and preference statistical features; 3)item features $\boldsymbol{x}^{i}$, such as item id, category, brand, and related statistic features; 4)context features $\boldsymbol{x}^{c}$, such as position and promotion information; 5)target item DR time series $\boldsymbol{S}^{d}$, which is first proposed in CVR-related works.
\\
\begin{figure*}[htbp]
\centering
\includegraphics[scale=0.58]{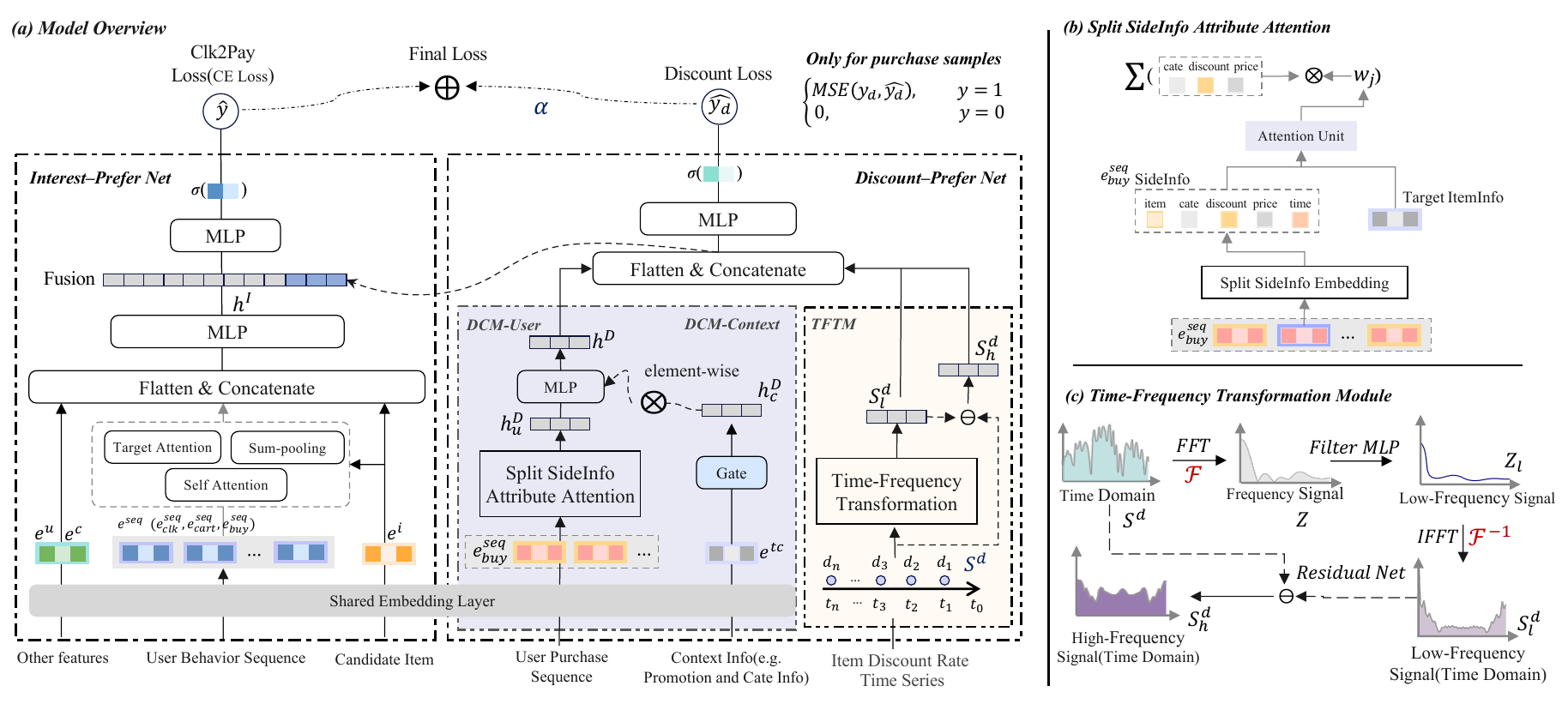}
\caption{Framework of the proposed Discount-Aware Network(DANet), which consists of the Interest-Prefer Net(IPN) and Discount-Prefer Net(DPN). The DPN includes three components: Time-Frequency Transformation Module(TFTM), Distribution Correction Module(DCM-User \& DCM-Context), and Supervised Regression Auxiliary Task.}
\label{fig:framework}
\end{figure*}

\subsection{Interest-Prefer Net}
As shown in Figure ~\ref{fig:framework}, the input features of the Interest-Prefer Net(IPN) are composed of non-sequential features (${x}^{u}$, ${x}^{i}$, ${x}^{c}$) and sequential features (${x}^{seq}$). After passing through the embedding layer, (${x}^{u}$, ${x}^{i}$, ${x}^{c}$) is assigned to a dense vector of lower dimensions (${e}^{u}$, ${e}^{i}$, ${e}^{c}$), where the numerical features are discretized into categorical features through the boundary values. The sequence features are constructed into an embedding of fixed dimensions through the sequence layer, e.g., ${e}^{seq}_{buy} =\left\{{e}_1^i, {e}_2^i, ..., {e}_L^i \right\}$, where ${e}_L^i$ denotes the item embedding of $ L $-th user purchase behavior, and $L$ denotes the fixed sequence length.
\par Specifically, we use Multi-Head Target-Attention(MHTA)~\cite{vaswani2017attention} on user different behavior sequences ${e}^{seq}$ to dynamically capture user interests. The query($Q$) is the candidate item ${e}^{i}$, and the historical behavior sequences ${e}^{seq}$ are taken as the key ($K$) and value ($V$). Then, ${e}^{u}$, ${e}^{i}$, ${e}^{c}$ and MHTA(${e}^{seq}$) are concatenated and fed into a Multi-Layer Perception (MLP), generating $h^I$ as the output of IPN:
\begin{equation}
h^I=f_i(\mathcal I(x))=f_i\left(\left[e^u ; e^i; e^c ; MHTA(e^{s e q})\right]\right)
\end{equation}
where $\mathcal I(x)$ denotes IPN, $[;]$ denotes the concatenation of embeddings, and $f_i(\cdot)$ is MLP output of IPN.
\subsection{Time-Frequency Transformation Module}
Given a DR time series $\boldsymbol{S}^{d} \in \mathbb{R}^{N }$ as follows:
\begin{equation}
S^d=[d_1,d_2,d_3...,d_N]
\end{equation}
where $N$ is the total number of time steps and $d_i$ denotes the average DR of all orders for the target item at the $ i $-th time step. In this work, we define a time step as one day and set $N$ to 400 to cover the past year, which is generally sufficient to capture the evolution of discounts for the target item.

To better characterize the temporal signal features of item DR, we employ a time-frequency analysis approach~\cite{ye2024frequency} to decompose the original signal into frequency components. Specifically, we first apply Fast Fourier Transform (FFT) to convert the time-domain signal into the frequency domain. Subsequently, a MLP $f_t(\cdot)$ is utilized for signal filtering, allowing to learn the influence of different frequency components while retaining the stable low-frequency information. Finally, Inverse Fast Fourier Transform (IFFT) is applied to transform the frequency domain signal back into the time domain, thereby modeling the long-term evolutionary trends of item DR.

\begin{equation}
Z = FFT(S^d),\ Z_{l} = f_t(Z),\ {S}^{d}_{l} = IFFT(Z_{l})
\end{equation}

The computational complexity in memory is $O(N)$, and we set the number of patterns N to the default value N = 400. The complexity of the full FFT transform is $(O(N log(N))$ ~\cite{zhou2022fedformerfrequencyenhanceddecomposed}.

Furthermore, we design a residual network to extract the non-stationary high-frequency signal component ${S}^{d}_{h}$. The process is expressed as:
\begin{equation}
{S}^{d}_{h} = {S}^{d}-{S}^{d}_{l}
\end{equation}

\subsection{Distribution Correction Module}
From Figure ~\ref{figure1}, we observe that DR is personalized and users have different sensitivities to DR for items in different categories. Besides, \cite{10486475} reflects that the DR ranges typically vary during different promotional periods. This section describes two correction networks: DCM-User and DCM-Context to model the above two biases. 

\par\textbf{DCM-User}: Purchase behavior is conducted at the end of the decision cycle and strongly reflects the DR preference of users.
Compared with click behavior, user purchase behavior is sparse and lacks temporal coherence. Therefore, we develop an Attribute-Attention mechanism based on sequence sideInfo splitting, which is used to model the user DR preferences.
\begin{equation}
\begin{aligned}
{h}^{D}_{u}=\mathcal A\mathcal A_{u}\left(E_A, {e}_1^i, {e}_2^i \cdots \cdot {e}_L^i\right) & =\sum_{j=1}^L a\left({e}_j^i, E_A\right)\left[{e}_{j d}^{i};{e}_{j c}^{i};{e}_{j p}^{i}\right]\qquad \\
& =\sum_{j=1}^L w_j\left[{e}_{j d}^{i};{e}_{j c}^{i};{e}_{j p}^{i}\right]
\end{aligned}
\end{equation}
where ${e}^{seq}_{buy}=\left\{{e}_1^i, {e}_2^i, ..., {e}_L^i \right\}$ is the list of embedding vectors of buy behavior of users with length of $L$, and ${e}_{j d}^{i}$, ${e}_{j c}^{i}$, ${e}_{j p}^{i}$ represents the discount, category, and price attribute vectors split from each vector in ${e}^{seq}_{buy}$, respectively. $E_A$ is the embedding vector of candidate item,  $a(\cdot)$ is a feed-forward network with output as the attention weight $w_j$. $\mathcal A\mathcal A_u(\cdot)$ as Attribute-Attention, which focuses on the DR, price, and category information of similar items in the buy behavior.

\par\textbf{DCM-Context}: Intuitively, the upper limit of DR could be different for different promotion periods. Inspired by the spirit of PEPNET~\cite{chang2023pepnetparameterembeddingpersonalized}, we dynamically combine bias embedding according to different promotion periods, which helps capture the cyclical DR bias brought by different promotion periods. Therefore, we select a set of contextual features to build a bias net related to platform time, and formalize the time period influencing factors into a bias embedding ${h}^{D}_{c}$, which is applied to ${h}^{D}_{u}$ and jointly form the output of the DCM ${h}^{D}$:


\begin{equation}
h^D=f_{d u}\left(h_u^D\right) \otimes {h_c^D} = f_{d u}\left(h_u^D\right) \otimes {GateNet}\left(e^{t c}\right)
\end{equation}
where $e^{tc}$ is the embedding of promotion context features. 
$GateNet(\cdot)$ as DCM-Context, and $f_{du}(\cdot)$ is an MLP  , which plays a role in dimension transformation.
\subsection{Supervised Regression Auxiliary Task}
By TFTM and DCM, we can obtain three DR vectors from three dimensions: user, item, and context. Then ${S}^{d}_{l}$, ${S}^{d}_{h}$ and ${h}^{D}$ are concatenated back to IPN as a supplementary feature for model learning. The widely-used CE loss is adopted as loss function to train IPN, $i.e.$,
\begin{equation}
\hat{y}=\sigma\left(f_m(\mathcal M(x))\right)=f_m\left(\left[h^I ; h^D ; {S}^{d}_{h} ; {S}^{d}_{l}\right]\right)
\end{equation}
\begin{equation}
Loss_m=-\frac{1}{|\mathcal{D}|} \sum_{(\boldsymbol{x}, y) \in \mathcal{D}}(y \log \hat{y}+(1-y) \log (1-\hat{y}))
\end{equation}
In addition, to make DPN produce a well-learned DR representation, we construct an explicit DR label ${y}_{d}$, use the mean square error (MSE) on purchase samples for supervised learning, and it is included in the final loss.
\begin{equation}
\hat{y}_d=\sigma\left(f_d(\mathcal {DPN}(x))\right)=f_d\left(\left[h^D ; {S}^{d}_{h} ; {S}^{d}_{l}\right]\right)
\end{equation}
\begin{equation}
\text {Loss}=\text {Loss}_{m}+\alpha * \begin{cases}\operatorname{MSE}\left(y_d, \hat{y}_d\right), & y=1 \\ 0, & y=0\end{cases}
\end{equation}
where $\hat{y}$ and $\hat{y}_{d}$ denotes the outputs of IPN and DPN, respectively, and $y$ is the CVR label. The variable $\alpha$ is a scaling hyperparameter whose optimal maximum value is determined by experiments. 


\section{Experiments}
\label{sec:experiments}
\subsection{Experimental Setup}

\subsubsection{\textbf{Datasets}}
To the best of our knowledge, there is no publicly available dataset for DR-related due to a lack of final payment price data. We collected and sampled the online service logs from Alibaba Tmall e-commerce platform between 2025/08/26 to 2025/11/25 to construct the experimental dataset $\mathcal{D}_{a}$. As shown in Table 1, the entire dataset is split into non-overlapped training dataset $\mathcal{D}_{train}$ (08/26-11/10) and testing set $\mathcal{D}_{test}$ (11/11-11/25).

\begin{table}[htbp]
    \centering
    \caption{Statistics of the established datasets.}
    \resizebox{\columnwidth}{!}{%
    \begin{tabular}{@{}lcccccc@{}}
        \toprule
        \#Dataset & \#Users & \#Items & \#Exposures & \#Clicks & \#Purchases \\ \midrule
        $\mathcal{D}_{train}$  & 9.76M  & 5.08M  & 2.41B  & 100.12M  & 5.08M      \\
        $\mathcal{D}_{test}$   & 4.45M  & 3.41M  & 0.96B  & 33.39M  & 1.67M      \\
        \bottomrule
    \end{tabular}
    \label{tab:datasets_stats}
    }
\end{table}

\subsubsection{\textbf{Evaluation Metrics}}
We adopt the area under the ROC curve (AUC) as the offline evaluation metric. pCVR(Post-click conversion rate) and GMV(Gross Merchandise Volume) are important monitoring indicators for online A/B testing, which are widely employed in industrial recommendation systems. 

\subsubsection{\textbf{Implementation Details}}
All experiments are implemented in distributed Tensorflow 1.4 and trained using 10 parameter servers and 4 Nvidia Tesla V100 16GB GPUs. Item ID has an embedding size of 64, category ID and brand ID have an embedding size of 32 while 8 for the other categorical features. We use an 8-head attention structure with a hidden size of 128 in IPN and DCM , and a 32-point FFT in TFTM. Training is performed using the Adagrad optimizer with a learning rate of 0.1 and a mini-batch size of 1024. 

\subsubsection{\textbf{Competitors}}
\textbf{PASBR}~\cite{feng2023modeling} employs lossless session graphs to model users' item price tolerance, introducing item price as a novel factor to influence preferences. \textbf{CoHHN}~\cite{Zhang_2022} adopts a dual-channel aggregation mechanism to simultaneously capture users' price and interest preferences through attention layers, resulting in enhanced CVR predictive performance compared to DNN. \textbf{DIN}~\cite{zhou2018deepnetworkclickthroughrate} addresses user interests about the target item, which has found substantial applications within the industry.

\subsection{Experimental Results}
\subsubsection{\textbf{SOTA}}: For offline evaluation, all experiments were repeated 5 times on $\mathcal{D}_{test}$. For online A/B testing, models are tested in turn in the our online app homepage. The experimental results are shown in Table 2. The main observations are summarized as follows: 1) The performance of the IPN-BASE model is basically the same as that of DIN, proving its effectiveness. 2) PASBR, which influences user preferences by introducing item prices and simulating price tolerance, becomes the runner-up method, further verifying the impact of price factors on purchase orientation. 3) CoHHN uses a dual-channel aggregation mechanism to replace the traditional MHTA in the item recommendation field, but the effect is not good. 4) Our DANet model achieves the best performance both offline and online and outperforms the runner-up approach by a large margin. 


\begin{table}[h]
    \centering
    \caption{Offline and online comparison results.}
    \resizebox{\columnwidth}{!}{%
    \begin{tabular}{lccc}
        \toprule
        Model & Offline & \multicolumn{2}{c}{Online} \\
        \cmidrule(lr){2-2} \cmidrule(lr){3-4}
        & AUC (mean±std.) & pCVR Gain & GMV Gain \\
        \midrule
        IPN-Base & 0.8345±0.00062 & 0\% & 0\% \\
        + DCM-User \small{(A)} & 0.8361±0.00131 & +0.85\% & +0.22\% \\
        + DCM-Context \small{(B)} & 0.8357±0.00276 & +0.59\% & +0.08\% \\
        + TFTM \small{(C)} & 0.8389±0.00052 & +2.65\% & +1.60\% \\
        + Auxiliary task \small{(D)} & 0.8347±0.00048 & +0.30\% & +0.02\% \\
        PASBR & 0.8370±0.00021 & +1.62\% & +0.79\% \\
        CoHHN & 0.8337±0.00253 & -0.74\% & -0.91\% \\
        DIN & 0.8343±0.00132 & -0.01\% & -0.04\% \\
        \textbf{DANet\small{(+A\&B\&C\&D)}} & \textbf{0.8479±0.00163} & \textbf{+3.63\%} & \textbf{+2.23\%} \\
        \bottomrule
    \end{tabular}
    }
\end{table}
\subsubsection{\textbf{Ablation Study}}: The order of improvement from large to small is TFTM>DCM-User>DCM-Context>Auxiliary task. 1) The significant improvement of the TFTM module means that the modeling paradigm of introducing the long-term evolution trends of item DR is conducive to the purchase-oriented learning of the model. 2) The improvement of the DCM-User and DCM-Context modules further shows that the item DR are affected by user and context bias, and personalized difference modeling is necessary. 3) Auxiliary task helps to learn better item DR representation for the final prediction by introducing explicit item DR labels. In addition, for online discount item PVR(Exposure Ratio), DANet improves \textbf{47.32\%} compared to IPN-Base, \textbf{7.56\% to 11.14\%}.
\subsubsection{\textbf{Case Study}}
\begin{figure}[htbp]
\centering
\includegraphics[scale=0.41]{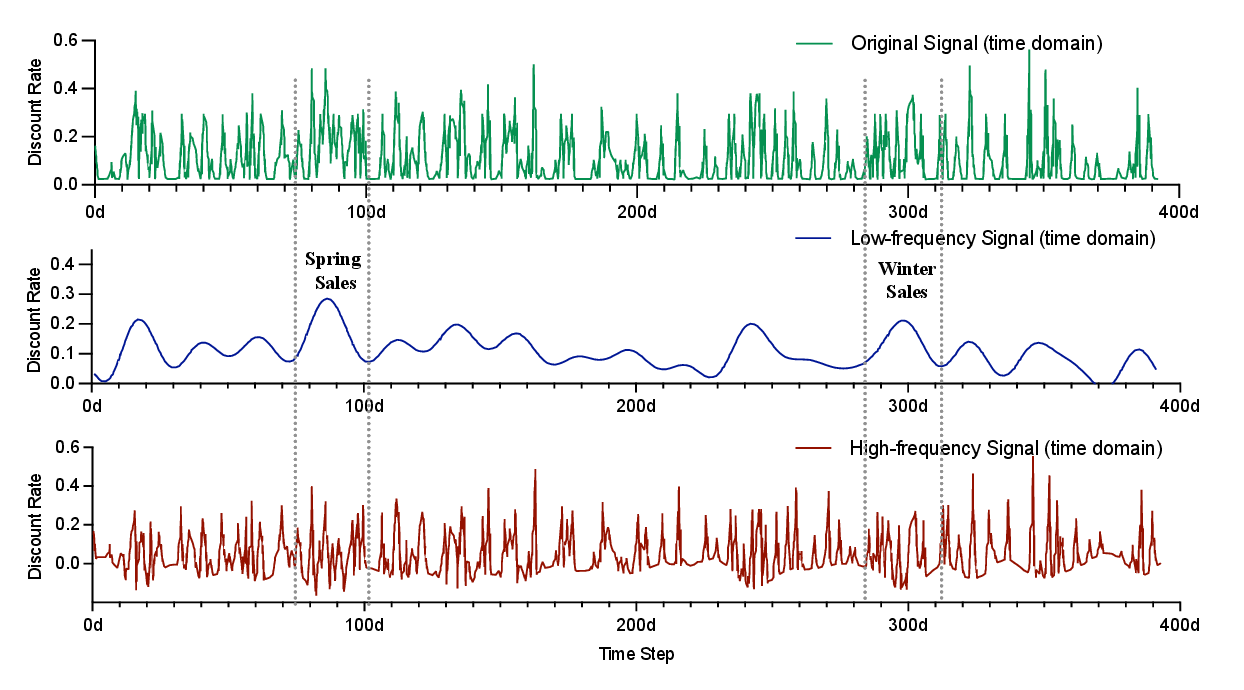}
\caption{Case analysis of ${S}^{d}$, ${S}^{d}_{l}$ and ${S}^{d}_{h}$.}
\label{fig:case}
\end{figure}

We evaluate the operational principle of TFTM by visualizing ${S}^{d}_{h}$ and ${S}^{d}_{l}$. By averaging over 1,000 representative samples, our results in Figure ~\ref{fig:case} confirm that TFTM effectively separates high- and low-frequency signals, where the low-frequency signal exhibits a smoother trend while the high-frequency signal contains more noise, aligning with our experimental predictions.
\label{subsec:exps}

\section{Conclusion} 
In this paper, we present a novel model named  Discount-Aware Network (DANet), which effectively integrates item DR into the prediction of CVR in online recommender systems. The inclusion of a time-frequency transformation module, a distribution de-bias module, and a supervised regression task enhances the model's accuracy and representation of item DR. Our experimental results validate the efficacy of DANet, demonstrating significant improvements in both offline and online performance metrics. These findings highlight the importance of incorporating discount dynamics into CVR modeling, paving the way for more effective personalized recommendations and better user engagement in e-commerce platforms.

\bibliographystyle{ACM-Reference-Format}
\balance
\bibliography{sample-base}


\end{document}